# Data Requirements for Evaluation of Personalization of Information Retrieval – A Position Paper

Nicholas J. Belkin[1], Daniel Hienert[2], Philipp Mayr[2], Chirag Shah[1]

[1]School of Communication & Information, Rutgers University, New Brunswick, NJ, USA
[2]GESIS, Cologne, Germany
belkin@rutgers.edu, Daniel.Hienert@gesis.org, Philipp.Mayr @gesis.org, chirags@rutgers.edu

**Abstract.** Two key, but usually ignored, issues for the evaluation of methods of personalization for information retrieval are: that such evaluation must be of a search session as a whole; and, that people, during the course of an information search session, engage in a variety of activities, intended to accomplish different goals or intentions. Taking serious account of these factors has major implications for not only evaluation methods and metrics, but also for the nature of the data that is necessary both for understanding and modeling information search, and for evaluation of personalized support for information retrieval (IR). In this position paper, we: present a model of IR demonstrating why these factors are important; identify some implications of accepting their validity; and, on the basis of a series of studies in interactive IR, identify some types of data concerning searcher and system behavior that we claim are, at least, necessary, if not necessarily sufficient, for meaningful evaluation of personalization of IR.

**Keywords:** Interactive IR, Information Seeking, Evaluation, Task, Session, User log data

## 1 Introduction

When people, seeking information in order to accomplish some task, or achieve some goal, engage with information retrieval (IR) systems, they often, perhaps most typically, conduct what one can term an *Information Seeking Session* (ISS). Although the minimal form of such a session can be a single query put to the system, a response by the system, a choice by the person of an item from the response, and the end of the session, both research in information behavior, and observation of behavior in operational systems, demonstrate that such behavior is not typical. Rather, an ISS often, perhaps most often, consists of a number of such iterations [1, 15, 19]. What, then, happens during the course of such an ISS? Although the typical IR system affords the person little more than the ability to formulate and reformulate queries, and to observe and select items from the system's response, in the form of links to information objects, there is substantial evidence that they are *intending* to accomplish many tasks or goals, in each such iteration, other than finding an information object that is relevant to the query that is submitted [2, 8, 13, 21]. Such goals may include, *inter alia*, learning about a domain, learning about the contents of a data base, comparing information objects, or identifying useful information objects through recognition, rather than specification. We term such goals *information seeking intentions* (not to be confused with the term *intent*, normally used to refer either to the general goal of the search as a whole [e.g. 3, 11] or the *topic* of the search [e.g. 18] ([22] uses the term *interactive intentions*).

Under this understanding of people's behaviors in interaction with IR systems, we, as have others [e.g. 7, 18], propose that IR is best construed as a sequence of interactions of the person with the IR system, motivated overall by some external task or goal, with each interaction being itself motivated by some information seeking intention. These interactions can be considered as sub-tasks that arise in the person's attempt to eventually achieve the overall goal of the ISS, which is to obtain that which is deemed useful in accomplishing the motivating task/goal. This view of IR has important implications for what it means to personalize support for IR interaction, and how to accomplish such personalization, for how to evaluate such support, and, importantly, for what data are required to both to accomplish such support, and to evaluate it. In this position paper, we address, in particular, the issue of the data required to accomplish and evaluate personalization of IR interaction.

## 2  Implications of the ISS Model of IR for Personalization and its Evaluation

It has been suggested that the ISS model of IR implies that *usefulness*, rather than relevance, is the appropriate criterion of evaluation of interactive IR [3, 6, 11]. This is based on the idea that support for the ISS should be evaluated with respect to the extent to which the entire ISS has been useful in helping the person to achieve the motivating task/goal. But, since the ISS consists of a sequence of information seeking intentions, evaluation must also be with respect to how useful the IR system has been in supporting these various intentions themselves, and with the usefulness of the support of those intentions to accomplishment of the motivating task/goal. Taking this stance suggests further that adaptation of the IR system to both motivating task/goal, and to the person's various information seeking intentions, are necessary to accomplish effective personalization of support in the ISS. This in turn suggests that, in order to accomplish, and evaluate the effectiveness of personalization, it is minimally necessary for the IR system to obtain data which will provide knowledge of the person's motivating task/goal, the goal of the ISS as a whole, and the goals of the various intentions. This leads to the question of just what data these might be, and how they might be collected.

In order to address the question just posed, we first examine just what data have already been collected in the course of the rather substantial record of research in the evaluation of interactive IR, and how it has been collected. We analyze and compare the data types collected in such studies in order to see if they can lead to the understandings which we propose are necessary to evaluate support for personalization in the manner we require. This is the subject of the next section of this paper. On the basis of these results, we then propose a research agenda, which could lead to meth-ods for identification and collection of the data necessary, beyond those which have already been identified, for proper evaluation of at least the aspect of personalization which we have described.

## 3   Existing Data Sets for IR and Interactive IR Evaluation

There exist a number of data sets for the evaluation of IR and Interactive IR (IIR) evaluation purposes. Table 1 gives an overview of a selection of existing data sets with their properties. We divided the data sets roughly in three different groups: (1) evaluation campaign data sets, (2) real-world data sets, and (3) lab study data sets. Evaluation campaign data sets normally involve a fixed corpus, given topics, queries and relevance judgements for all or a set of result documents. The goal here is to optimize the ranking function of the system based on given topics and the expected relevant results. Thereby the focus of investigation has opened from the query level (TREC Web Track [7]), over the session (TREC Dynamic Track [23]) to the task (TREC Task Track [20]). The TREC session track [5] in contrast combines a given corpus, topics, queries and relevance judgments with retrieved results, click data and dwell times from crowd workers conducting searches within a session. Similar is the INEX Interactive Track [14] where different tasks of real users are conducted. The current PIR-CLEF[1] initiative then moves the scope again to tasks and adds some personal information about the user and a number of weighted terms describing the user interests based on documents of interest.

A second group of evaluation data sets comes from real-world search engines. For web search there exist e.g. the Yandex Web Search click data [17] which includes queries, retrieved results, click data and document dwell times extracted from transaction logs. Document relevance is computed from dwell times on documents. For discipline specific search there exist e.g. the SUSS data set with retrieval sessions performed in a social sciences academic search engine [12]. These log data sets normally contain a high number of search sessions and a variety of users which make them appropriate for large-scale analysis.

Evaluation data sets from lab studies examine how real users conduct a certain task type with a given topic. From the data side the desire is to log as much information as possible in order to analyze user behavior from this data. This includes logging all interaction with the system including keyboard and mouse interaction and in addition the eye movements. For subjective measures subjects are interviewed before and after the search session e.g. to learn something about the task's difficulty and success. Additionally, to identify learning steps and decisions within the search session and certain query segments, a number of user interviews have to be conducted after the search session. Given a task, the documents usefulness needs to be assessed in relation to the overall task or sub task (not to the query). Another important issue is to understand the role of each query segment for the overall task. A current investigation is to ask the user for the intention of each individual query segment. The question can be: Was it to find new information or to evaluate already found information objects?

---

[1] http://www.ir.disco.unimib.it/pirclef2017/description-of-the-laboratory/

**Table 1.** Overview of properties of different (I)IR evaluation data sets.

| Dataset | Artificial vs. Real-World | Role of users | Focus | Size | Fixed corpus | Topics | Queries | Relevance Judgements | Retrieved Results | Click data | Dwell times | Tasks | User Actions | Keyboard/ Mouse Data | Eye-Tracking data | Document usefulness to task | Pre-, Post-Questionnaires | Query segment information, e.g. intentions, reformulations |
|---|---|---|---|---|---|---|---|---|---|---|---|---|---|---|---|---|---|---|
| TREC Web Track/Core Track [7] | Artificial | Inherent in Topics & Judgements | Query | 50 topics | ✓ | ✓ | ✓ | ✓ | | | | | | | | | | |
| TREC Dynamic Track [23] | Artificial | Inherent in Topics & Judgements | Session | 53 topics | ✓ | ✓ | ✓ | ✓ | | | | | | | | | | |
| TREC Task Track [20] | Artificial | Inherent in tasks extracted from logs | Task | 50 queries | ✓ | ✓ | ✓ | ✓ | | | | ✓ | | | | ✓ | | |
| TREC Session Track [5] | Controlled | Crowd-workers perform search sessions | Session | 1,257 sessions | ✓ | ✓ | ✓ | ✓ | ✓ | ✓ | ✓ | | | | | | | |
| INEX Interactive Track [14] | Controlled | Users conduct tasks | Task | 7 Tasks | ✓ | | ✓ | ✓ | ✓ | ✓ | ✓ | ✓ | ✓ | | | | | ✓ |
| PIR-CLEF | Controlled | Users conduct tasks | Task | 10 sessions | ✓ | ✓ | ✓ | ✓ | ✓ | ✓ | ✓ | ✓ | ✓ | | | | | |
| Yandex Web Search [17] | Real | Logged users | Session | ~34M sessions | | | ✓ | | ✓ | ✓ | ✓ | ✓ | | | | | | |
| SUSS [12] | Real | Logged users | Session | ~500K sessions | | | ✓ | | | ✓ | ✓ | ✓ | | ✓ | | | | |
| Writing Task Data Set [10] | Controlled | Users conduct tasks | Task | 150 topics | ✓ | ✓ | ✓ | | | ✓ | ✓ | ✓ | ✓ | | | | | |
| Recent IIR Lab Studies (e.g. [13]) | Controlled | Users conduct tasks | Task | 80 tasks | | | ✓ | | ✓ | ✓ | ✓ | ✓ | ✓ | ✓ | ✓ | ✓ | ✓ | ✓ |

## 4 Research Agenda

In reviewing Table 1, we note several important differences between the different data sets, which are especially significant for evaluation of personalization. The most obvious is that only one data set includes detailed information about the search session at the query segment level (last row of Table 1). Since data at this level would be crucial for evaluation of the aspect of personalization we discuss, it's clear that the methods used in this, and similar studies, need to be considered for evaluation of personalization. However, this type of study suffers from at least two other problems: a small number of cases (users, tasks, topics); and, controlled, rather than real, tasks and task types. A conclusion that one can draw from this comparison, is that what is required is some means for incorporating, in one general type of study, methods which allow: the collection of (relatively) large numbers of cases, of real tasks, addressed over whole search sessions, segmented and identified by information search intentions. Developing a means for doing this, effectively defines a research agenda for the design of studies which aim to evaluate personalization of support for IR. Below, we provide some examples of the types of data that would need to be collected in such studies, as contemplated for, e.g. PIR-CLEF 2018.

As an example, we could extract various aspects of learning that take place throughout the search. Specifically, we should try to understand how the searcher is learning about the task and the domain as he/she retrieves and assesses information, and how that learning affects his/her ongoing search activities. Some of the questions to ask the searcher or an assessor for eliciting such information are:

- What has been learned from the domain knowledge?
- What has been learned from the content obtained?
- How useful is what has been learned for the sub task?
- How well did the system support learning?

Another important aspect that we find useful to elicit from an IIR study is that of evaluation. Specifically, we believe it is important to discover how *searchers* (not external judges) evaluate an item for correctness and usefulness, and not just relevance. Searchers are also often comparing several relevant/useful items and picking the best and it would be interesting to know how they make such decisions. Some of the questions that could be asked to the searcher or an assessor to gather such information are:

- Which items were evaluated?
- What was evaluated?
- What were the criteria?
- How useful were the items for the sub task?
- How well did the system in supporting evaluating the items?

These two types of intention of course do not cover all information seeking intentions that could occur during an ISS (see, e.g. [13] or [21] for more inclusive lists), but, as examples, they indicate the nature of the data that would be required to evaluate personalization to any such intention. We hope that these examples are sufficient to indicate at least some aspects of what would need to be covered in a research agenda for specification of data types for evaluation of personalization of information retrieval.

## 5    Conclusion

We have proposed a view of IR that implies that personalization should be with respect not only to context, but to the various information search intentions that people have during the course of an information seeking session. We have identified some types of data which we claim would be necessary in order to evaluate the effectiveness of such personalization. We suggest that, learning just what data are necessary, and developing methods to gather such data, constitute the basis for a research agenda central to the general task of evaluation of personalization of support for IR. This could also be a starting point for considering the nature of the task for PIR-CLEF 2018.

## 6    Acknowledgments

This work was partly funded by Deutsche Forschungsgemeinschaft (DFG), grant no. MA 3964/5-1; the AMUR project at GESIS; and, by the National Science Foundation, grant no. IIS-1423239.

## 7    References

1. Bates, M.J.: The design of browsing and berrypicking techniques for the online search interface. Online Rev. 13, 5, 407–424 (1989).
2. Belkin, N.J.: Intelligent information retrieval: Whose intelligence? In: ISI '96: Proceedings of the Fifth International Symposium for Information Science. pp. 25–31 (1996).
3. Belkin, N.J.: On the evaluation of interactive information retrieval systems. In: Larsen, B. et al. (eds.) The Janus Faced Scholar. A Festschrift in Honour of Peter Ingwersen. pp. 13–21 , Copenhagen: Royal School of Library and Information Science (2010).
4. Broder, A.: A Taxonomy of Web Search. SIGIR Forum. 36, 2, 3–10 (2002).
5. Carterette, B. et al.: Overview of the TREC 2014 session track. Proceedings of TREC 2014, (2014).
6. Cole, M. et al.: Usefulness as the criterion for evaluation of interactive information retrieval. In: Proceedings of the Workshop on Human-Computer Interaction and Information Retrieval. pp. 1–4 (2009).
7. Collins-Thompson, K. et al.: TREC 2014 web track overview. Proceedings of TREC 2014, (2014).
8. Cool, C., Belkin, N.J.: A Classification of Interactions with Information. In: Bruce, H. et al. (eds.) Emerging frameworks and methods. Proceedings of the Fourth International Confer-ence on Conceptions of Library and Information Science (CoLIS4). pp. 1–15 Libraries Unlimited, Greenwood Village, CO (2004).

9. Fuhr, N.: A Probability Ranking Principle for Interactive Information Retrieval. Inf Retr. 11, 3, 251–265 (2008).
10. Hagen, M. et al.: How Writers Search: Analyzing the Search and Writing Logs of Non-fictional Essays. In: Kelly, D. et al. (eds.) Proceedings of the 1st ACM SIGIR Conference on Human Information Interaction and Retrieval (CHIIR 16). pp. 193–202 ACM (2016).
11. Hienert, D., Mutschke, P.: A Usefulness-based Approach for Measuring the Local and Global Effect of IIR Services. In: Proceedings of the 2016 ACM on Conference on Human Information Interaction and Retrieval. pp. 153–162 ACM, New York, NY, USA (2016).
12. Mayr, P., Kacem, A.: A Complete Year of User Retrieval Sessions in a Social Sciences Academic Search Engine. In: 21st International Conference on Theory and Practice of Digital Libraries (TPDL 2017). (2017).
13. Mitsui, M. et al.: Extracting Information Seeking Intentions for Web Search Sessions. In: Proceedings of the 39th International ACM SIGIR Conference on Research and Development in Information Retrieval. pp. 841–844 ACM, New York, NY, USA (2016).
14. Pharo, N. et al.: Overview of the INEX 2010 Interactive Track. In: Proceedings of the 9th International Conference on Initiative for the Evaluation of XML Retrieval: Comparative Evaluation of Focused Retrieval. pp. 227–235 Springer-Verlag, Berlin, Heidelberg (2011).
15. Rieh, S.Y., Xie, H.: Analysis of Multiple Query Reformulations on the Web: The Interactive Information Retrieval Context. Inf Process Manage. 42, 3, 751–768 (2006).
16. Rose, D.E., Levinson, D.: Understanding User Goals in Web Search. In: Proceedings of the 13th International Conference on World Wide Web. pp. 13–19 ACM, New York, NY, USA (2004).
17. Serdyukov, P. et al.: WSCD2013: Workshop on Web Search Click Data 2013. In: Proceedings of the Sixth ACM International Conference on Web Search and Data Mining. pp. 787–788 ACM, New York, NY, USA (2013).
18. Teevan, J. et al.: Personalizing Search via Automated Analysis of Interests and Activities. In: Proceedings of the 28th Annual International ACM SIGIR Conference on Research and Development in Information Retrieval. pp. 449–456 ACM, New York, NY, USA (2005).
19. Teevan, J. et al.: Potential for Personalization. ACM Trans Comput-Hum Interact. 17, 1, 4:1–4:31 (2010).
20. Verma, M. et al.: Overview of the TREC Tasks Track 2016. In: Proceedings of TREC 2016. (2016).
21. Xie, H.I.: Shifts of Interactive Intentions and Information-seeking Strategies in Interactive Information Retrieval. J Am Soc Inf Sci. 51, 9, 841–857 (2000).
22. Xie, I.: Interactive Information Retrieval in Digital Environments. IGI Global, Hershey, PA, USA (2008).
23. Yang, G.H., Soboroff, I.: TREC 2016 Dynamic Domain Track Overview. In: Proceedings of TREC 2016. (2016).